\def\chandra{{\it Chandra\/}}

\def\hst{{\it {\it HST}\/}}

\def\xeus{{\it XEUS\/}}

\def\ltsima{$\; \buildrel < \over \sim \;$}
\def\simlt{\lower.5ex\hbox{\ltsima}}
\def\gtsima{$\; \buildrel > \over \sim \;$}
\def\simgt{\lower.5ex\hbox{\gtsima}}
        
\documentclass[preprint]{aastex}

\begin{document}


\title{The Chandra Deep Field North Survey. VII. X-ray Emission from Lyman Break Galaxies}


\author{W.N.~Brandt,$^1$ 
A.E.~Hornschemeier,$^1$ 
D.P.~Schneider,$^1$
D.M.~Alexander,$^1$ 
F.E.~Bauer,$^1$ 
G.P.~Garmire,$^1$ and  
C.~Vignali$^1$}

\footnotetext[1]{Department of Astronomy \& Astrophysics, 525 Davey Laboratory, 
The Pennsylvania State University, University Park, PA 16802}


\begin{abstract}
We present results from stacking analyses, using the 1~Ms \chandra\ Deep Field North
data, that constrain the X-ray emission of Lyman break galaxies at $z\approx$~2--4.  
Stacking the counts from 24 individually undetected 
Lyman break galaxies located within the Hubble Deep Field
North, we have obtained average detections of these objects in the resulting
0.5--8.0~keV and 0.5--2.0~keV images; these images have effective exposure times of 
22.4~Ms (260 days). Monte Carlo testing empirically shows
the detections to be highly significant. The average rest-frame 2--8~keV luminosity 
of a Lyman break galaxy is derived to be $\approx 3.2\times 10^{41}$~erg~s$^{-1}$,
comparable to that of the most X-ray luminous starbursts in the local Universe.
The observed ratio of X-ray to $B$-band luminosity is somewhat, but not
greatly, higher than that seen from local starbursts. The X-ray emission probably 
arises from a combination of high-mass X-ray binaries, ``super-Eddington''
X-ray sources, and low-luminosity active galactic nuclei. 
\end{abstract}




\section{Introduction}

Over the last few years, the Lyman break technique has been used extensively to
isolate galaxies at $z\approx$~2--4 (e.g., Steidel et~al. 1996; 
Lowenthal et~al. 1997; Dickinson 1998; Steidel et~al. 1999) observed near 
the peak in the cosmic star-formation rate (e.g., Blain et~al. 1999). 
These objects often exhibit stellar and interstellar absorption lines
characteristic of local starburst galaxies. Their 
morphologies are varied, with multiple knots of emission
and diffuse wispy tails that suggest nonrelaxed systems. The Lyman break
technique has found particular application in the Hubble Deep Field North
(HDF-N; Williams et~al. 1996) due to the excellent imaging data 
and photometry available there, although the small angular size of 
the HDF-N limits somewhat its utility for statistical studies 
(e.g., Dickinson 1998).  

We have recently completed an $\approx 1$~Ms observation of the HDF-N and its
environs with the {\it Chandra X-ray Observatory\/} (Weisskopf et~al. 2000): 
the \chandra\ Deep Field North (\hbox{CDF-N}; Brandt et~al. 2001b, hereafter
Paper~V). In addition to probing luminous active galactic nuclei (AGN) 
throughout the Universe, this survey is also useful for studying
X-ray emission from more ``normal'' galaxies where the emission originates
from X-ray binaries, supernova remnants, and low-luminosity AGN (LLAGN; 
e.g., Hornschemeier et~al. 2001, Paper~II; Brandt et~al. 2001a, Paper~IV). 
Stacking techniques are particularly powerful in this regard, 
since they allow detection of X-ray emission from objects lying below the detection
limit for individual sources. In Paper~IV we used stacking techniques
to constrain the X-ray emission from normal spiral galaxies out to
$z\approx$~0.5--1, finding that their X-ray luminosities were not
more than a factor of $\approx 2$ larger (per unit $B$-band luminosity)
than those of spiral galaxies in the local Universe ($z<0.01$). 
Stacking analyses using the 480~ks of data available at the time 
allowed effective exposures of 5--8~Ms to be achieved. 

With the additional \chandra\ data recently obtained and improvements in 
our data processing techniques, we are now able to extend our
constraints on galaxy X-ray luminosities to higher redshift. This
is of cosmological interest since the X-ray luminosities of normal galaxies
are expected to evolve with redshift due to the changing cosmic
star-formation rate (e.g., van~Paradijs 1978; White \& Ghosh 1998;
Ghosh \& White 2001; Ptak et~al. 2001). Here we present constraints 
on the X-ray emission from the Lyman break galaxy population found 
in the HDF-N itself. 

The Galactic column density along this line of sight
is $(1.6\pm 0.4)\times 10^{20}$~cm$^{-2}$ (Stark et~al. 1992).
$H_0=65$~km~s$^{-1}$ Mpc$^{-1}$, 
$\Omega_{\rm M}=1/3$, and 
$\Omega_{\Lambda}=2/3$
are adopted throughout this paper. 
Coordinates throughout this paper are J2000.


\section{Data, Analysis, and Results}

We have used the Advanced CCD Imaging Spectrometer 
(ACIS; G.P. Garmire et~al., in preparation)
data sets and basic analysis methods described in 
Paper~V. We employ three standard X-ray bands: 0.5--8.0~keV (full band), 
0.5--2.0~keV (soft band), and 2--8~keV (hard band). We use 
data screened with the restricted ACIS grade set defined in 
Table~2 of Paper~V, since this screening provides significantly
improved signal-to-noise for faint sources. We restrict our 
focus to the data obtained within the HDF-N itself; our X-ray
coverage is deepest in this area, and the \chandra\ background 
and point-spread function (PSF) are relatively uniform over the 
HDF-N. \chandra\ positions within the HDF-N are good to
within $0\farcs 6$. 

For the stacking analysis, we selected galaxies 
in the HDF-N itself using the spectroscopic redshift catalogs 
of Cohen et~al. (2000), Cohen (2001), 
Dawson et~al. (2001), and references therein. We considered 
the $z=$~2--4 galaxies spectroscopically identified in the 
HDF-N; there are 28 such objects, and most 
were found using the Lyman break technique (the \hst\ 
$U_{300}$ filter has allowed Lyman break galaxies to be 
found down to $z\approx 2$ in the HDF-N; e.g., Dickinson 1998).  
Prior to stacking, we verified that none of the objects being stacked
was individually detected by \chandra. This was accomplished by 
searching for nearby (within $2\arcsec$) X-ray sources found by 
{\sc wavdetect} with a false-positive probability threshold of 
$10^{-5}$ (Dobrzycki et~al. 1999; Freeman et~al. 2001); it
led to the removal of one source, the $z=3.479$ AGN
123639.6+621230 (associated with 
CXOHDFN~J123639.5+621230). We have also manually 
inspected \chandra\ images in all three of the
standard X-ray bands and removed 
three Lyman break galaxies lying near unrelated X-ray sources.
123645.3+621153 was removed because it lies within the region
covered by the faint extended source CXOHDFN~J123645.0+621142. 
123645.8+621411 and 123654.7+621314 were 
removed because they lie relatively
near (6--8$\arcsec$) the bright X-ray sources 
CXOHDFN~J123646.3+621404 and CXOHDFN~J123655.4+621311, 
respectively, and could be affected by the low-level 
wings of the \chandra\ PSF. While we consider it
unlikely that these three Lyman break galaxies suffer from 
significant contamination by the unrelated X-ray sources, 
we wanted to be cautious and avoid any problems with the
background estimation. 
Figure~1 shows the redshifts and $R$ magnitudes of the 24 
selected objects. The median redshift of the selected objects 
is $z=2.92$, corresponding to a median lookback time of
11.9~Gyr (84\% of the age of the Universe) and a median
luminosity distance of 25,800~Mpc (Hogg 1999). 
These galaxies have a range of morphologies but generally appear 
nonrelaxed. 
They do not show signs of containing AGN in their optical spectra,
although the spectral quality is often limited. They were not
identified as AGN candidates by Jarvis \& MacAlpine (1998), 
Conti et~al. (1999), or Sarajedini et~al. (2000). Two of
the 24 galaxies have weak (6--8~$\mu$Jy) radio emission at
8.5~GHz (Richards et~al. 1998); this emission could be due
to either starburst activity or AGN. 

For each of the standard X-ray bands, we stacked ``cutout''
images centered on the positions of the 24 selected galaxies. 
For the photometry below we use an approximately circular
aperture comprised of the 30 pixels with 
centers within $1\farcs 5$ of the stacking
position (see \S3.4 of Paper~IV). In Figure~2 we show 
the number of full-band and soft-band counts 
obtained for each of the individual galaxies being stacked, 
and in Figure~3 we show images created from the stacked data. 
The stacked full-band image has an effective exposure time of 
22.44~Ms (260~days). In this image, we find 164 counts 
within the 30-pixel aperture, while 134.3 are expected
from the background. The Poisson probability of obtaining 164 counts 
or more when 134.3 are expected is $5\times 10^{-3}$, indicating 
a detection at the 99.5\% confidence level. 
In the stacked soft-band image (22.46~Ms exposure), we find 43 counts
when 26.2 are expected from background, corresponding to a detection
at the 99.9\% confidence level. The higher significance of the soft-band
detection is most likely due to the lower background in this band. 
No significant detection is obtained in the hard band; unfortunately, 
the spectral constraint implied by the hard-band nondetection is
not physically interesting.  

Following Appendix~A of Paper~IV, we performed Monte Carlo stacking
simulations to assess false-detection probabilities empirically. 
For each band, we performed 100,000 trials where 
we stacked 24 random positions using the same photometry aperture as 
was used for the Lyman break galaxies above. The random positions 
were chosen to lie within $15\arcsec\times 15\arcsec$ 
``local background regions'' centered on each of the Lyman break 
galaxies (avoiding known \chandra\ sources) to reproduce the actual 
background as closely as possible.
Figure~4 displays the number of trials giving a particular number of 
counts for the full and soft bands. The resulting distributions 
are very nearly Gaussian. 
In the full band 518 of the 100,000 stacking trials gave 164 counts
or more; the corresponding detection confidence level of 99.5\%
is in excellent agreement with that computed from Poisson statistics 
above. In the soft band 103 of the 100,000 stacking trials gave 43 counts
or more; again the corresponding detection confidence level of 99.9\%
agrees well with Poisson statistics. 
Again, no significant detection is obtained in the hard band. 

We performed several additional tests to verify the robustness
of our results. 
If we include the three Lyman break galaxies excluded above
because they lie near to unrelated X-ray sources (123645.3+621153,
123645.8+621411, and 123654.7+621314), the statistical significance
of our results is improved. 
We varied the size of the local background regions to examine if 
background gradients could be confusing the stacking analysis; our 
results remain highly significant for local background region sizes of  
$5\arcsec\times 5\arcsec$, 
$9\arcsec\times 9\arcsec$, and 
$21\arcsec\times 21\arcsec$. 
Because use of the local background regions limits the number of 
independent random positions that can be chosen in the Monte Carlo 
simulations, we also performed simulations with 
``global background regions.'' For these 
simulations we drew the random positions from larger areas, 
not restricting the positions to lie close to the Lyman break galaxies. 
The statistical significance of our results remained the same or was
slightly improved. 
Finally, we varied the size of our photometry aperture, and our
results remain significant for reasonable aperture-size choices.
In fact, reducing the aperture size to include only the 14 pixels
with centers within $1\farcs 0$ of the stacking position improves
the statistical significance by a further factor of $\approx 5$
in both the full and soft bands. 

We note that our results cannot be explained 
by instrumental effects (e.g., subtle cosmic ray afterglows) 
since such effects will not be preferentially correlated with 
Lyman break galaxies. Moreover, unrelated galaxies at lower redshift 
that by chance lie within our stacking apertures cannot explain the 
observed signal since these galaxies will be found equally in both
the Lyman break galaxy stacking and the Monte Carlo simulations. 


\section{Discussion}

The stacking analyses above extend our ability to detect
relatively normal galaxies in the X-ray band to substantially
higher redshifts ($z=$~2--4) than 
was previously possible. At the median
redshift of our sample of $z=2.92$, the observed-frame full and
soft bands correspond to rest-frame bands of 2.0--31.4~keV and
2.0--7.8~keV, respectively. Provided the basic X-ray production
mechanisms of Lyman break galaxies are similar to those of spiral
galaxies in the local Universe, the emission in these bands
should have an effective power-law photon index of $\Gamma\approx 2$
(e.g., Kim, Fabbiano, \& Trinchieri 1992; Ptak et~al. 1999); 
this is a typical photon index for both X-ray binaries and LLAGN. 
In the soft band, where we obtain the highest significance 
detection, the measured count rate 
($7.5\times 10^{-7}$~count~s$^{-1}$) corresponds to an average
observed-frame flux of $4.0\times 10^{-18}$~erg~cm$^{-2}$~s$^{-1}$
and an average rest-frame 2.0--7.8~keV luminosity of 
$3.2\times 10^{41}$~erg~s$^{-1}$ (at the median 
redshift).\footnote{For comparison, our soft-band flux
limit for a single source is 
$\approx 3.0\times 10^{-17}$~erg~cm$^{-2}$~s$^{-1}$, 
corresponding to a rest-frame 2.0--7.8~keV luminosity 
of $\approx 2.4\times 10^{42}$~erg~s$^{-1}$.}
Note that these Lyman break galaxies are $\approx 30$ times more
luminous from 2--8~keV than the 11 normal spiral galaxies 
at $z\approx 0.5$ studied in \S4.2 of Paper~IV. 
Since these Lyman break galaxies are thought to contain
significant starburst activity (see \S1), it is appropriate
to compare their X-ray luminosities to those of local starbursts. 
On average the 2--8~keV luminosities of these Lyman break galaxies 
are $\approx 5$ times larger than that of the starburst galaxy M82 
(e.g., Griffiths et~al. 2000; Kaaret et~al. 2001)
and are comparable to that
of NGC~3256, one of the most X-ray luminous starbursts in the 
local Universe (e.g., Moran, Lehnert, \& Helfand 1999; 
Lira et~al. 2001). The star-formation rates in these Lyman
break galaxies are uncertain but are thought to be often
$\simgt$~30--50~M$_\odot$~yr$^{-1}$, compared to 
$\approx 10$~M$_\odot$~yr$^{-1}$ for M82 and
$\approx 40$~M$_\odot$~yr$^{-1}$ for NGC~3256. 

The HDF-N Lyman break galaxies are optically 
luminous with $B$-band luminosities somewhat greater than 
present-day $L^\ast$; a present-day $L^\ast$ galaxy has a 
$B$-band luminosity density of 
$l_{\rm B}^\ast=1.0\times 10^{29}$~erg~s$^{-1}$~Hz$^{-1}$
(e.g., Blanton et~al. 2001). 
The 24 Lyman break galaxies used in the stacking 
analysis above have a mean rest-frame 
$l_{\rm B}=2.4\times 10^{29}$~erg~s$^{-1}$~Hz$^{-1}$ [the
median $l_{\rm B}=1.7\times 10^{29}$~erg~s$^{-1}$~Hz$^{-1}$, 
and the range is 
$l_{\rm B}=$~(0.4--6.6)$\times 10^{29}$~erg~s$^{-1}$~Hz$^{-1}$]; 
this is $\approx 1.3$ times larger than for the 11 normal spiral 
galaxies at $z\approx 0.5$ studied in \S4.2 of 
Paper~IV.\footnote{The rest-frame $l_{\rm B}$ values for the 
Lyman break galaxies have been calculated 
using the observed-frame $H$-band data in Table~2 of 
Papovich, Dickinson, \& Ferguson (2001) and the irregular
galaxy spectral energy distribution from 
Coleman, Wu, \& Weedman (1980). The good match between the
$H$-band data and rest-frame $B$ for these objects removes
nearly all of the dependence of $l_{\rm B}$ on the 
spectral energy distribution.} Therefore, the 
HDF-N Lyman break galaxies are 
$\approx (30/1.3)\approx 23$ 
times more X-ray luminous
per unit $B$-band luminosity than normal spiral galaxies at 
$z\approx 0.5$ (also see Shapley, Fabbiano, \& Eskridge 2001). 
An elevated ratio of X-ray to $B$-band luminosity is 
observed in many starburst galaxies at low redshift. 
The starburst galaxies in \S3 of David, Jones, \& Forman (1992) have 
X-ray to $B$-band luminosity ratios $\approx$~3--15 times higher 
than for normal galaxies (for example, this ratio is 
$\approx 13$ times higher for M82). Thus, the X-ray to $B$-band 
luminosity ratios for Lyman break galaxies are somewhat, but
not greatly, higher than those of local starbursts. If the
Lyman break galaxies suffered from more dust obscuration than local 
starbursts, this could explain their higher X-ray to $B$-band 
luminosity ratios; dust would attenuate $B$-band emission 
substantially more than 2--8~keV emission. 

Theoretical efforts to predict the X-ray evolution of galaxies
(e.g., van~Paradijs 1978; White \& Ghosh 1998; Ghosh \& White 2001; 
Ptak et~al. 2001) have typically focused 
on their low-mass X-ray binary (LMXB) populations which are
predicted to peak in luminosity at $z\approx 1.5$. Much less
work has been carried out, however, at the higher redshifts under study here. 
If the X-ray emission of Lyman break galaxies is indeed produced
via the same mechanisms as those operating in local starburst
galaxies, then high-mass X-ray binaries (HMXBs) and ``super-Eddington''
X-ray sources (e.g., Makishima et~al. 2000) are likely 
to produce much of the observed emission in the rest-frame
2--8~keV band. Unlike LMXBs, which have an evolutionary 
timescale of $\sim 10^9$~yr, the X-ray emission from HMXBs
should track the cosmic star-formation rate closely (any LMXBs
contributing to the observed X-ray emission would have formed
at $z\sim 5$). Given the results from submillimeter and other surveys 
which are sensitive to star formation in dusty environments, cosmic 
star formation is likely to peak at roughly the redshifts under study 
here (e.g., Blain et~al. 1999). The cosmic evolution
of super-Eddington X-ray sources 
is impossible to predict reliably at present given that even
the basic nature of these objects is poorly understood at low
redshift. LLAGN with 2--8~keV luminosities
$\simlt 2.4\times 10^{42}$~erg~s$^{-1}$
could also contribute to the observed X-ray emission
from some of these Lyman break galaxies, although none of the
galaxies under study here has been found to contain an LLAGN (see \S2); 
again, even the most basic properties of LLAGN at these redshifts 
are poorly constrained by current data. 

Finally, these results show that the \chandra\ ACIS performs well at 
source detection even with effective exposure times of 260 days
(the current longest \chandra\ exposures are 11.6~days for the
CDF-N and the \chandra\ Deep Field South). Any systematic effects
that cause the sensitivity to deviate from that expected due
to photon statistics appear mild. Stacking analyses 
using deeper observations with \chandra\ should allow this 
work to be extended. Missions such as \xeus\ should ultimately 
allow the X-ray study of individual Lyman break galaxies. 


\acknowledgments

This work would not have been possible without the enormous efforts 
of the entire \chandra\ and ACIS teams. 
We thank 
O.~Almaini, 
A.C.~Fabian,
E.C.~Moran,  
A.~Ptak, 
A.E.~Shapley,
and an anonymous referee for helpful discussions. 
We gratefully acknowledge the financial support of
NASA grant NAS~8-38252 (GPG, PI),
NSF CAREER award AST-9983783 (WNB, DMA, FEB),  
NASA GSRP grant NGT~5-50247 and the Pennsylvania Space Grant Consortium (AEH), 
NSF grant AST-9900703~(DPS), and 
NASA LTSA grant NAG5-8107 (CV). 


\clearpage


\begin{figure}
\epsscale{0.8}
\figurenum{1}
\plotone{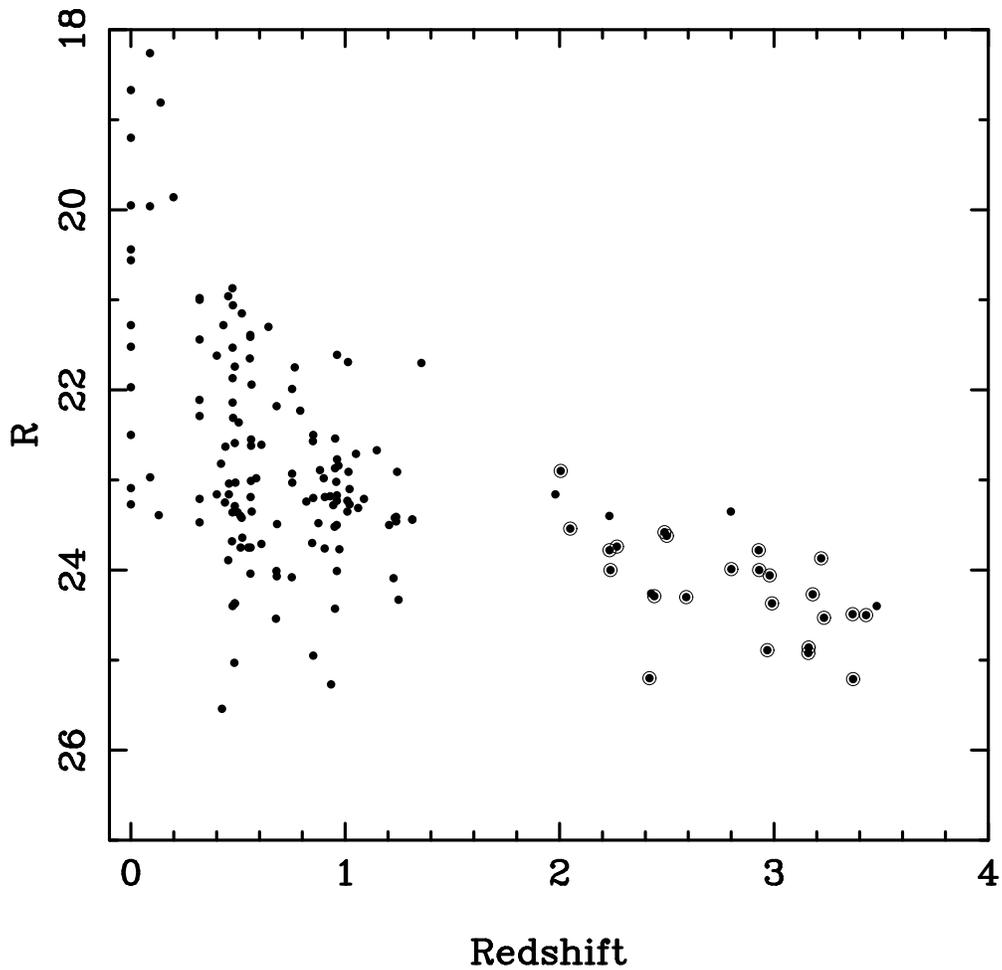}
\vspace{-0.1truein}
\caption{$R$ magnitude versus redshift for the HDF-N galaxies 
with spectroscopic redshifts in Cohen et~al. (2000), Cohen (2001), 
Dawson et~al. (2001), and references therein. The circled galaxies 
are those used in our stacking analysis. The four objects from 
$z=$~2--4 that are not circled are discussed in \S2.}
\end{figure}


\begin{figure}
\epsscale{0.8}
\figurenum{2}
\plotone{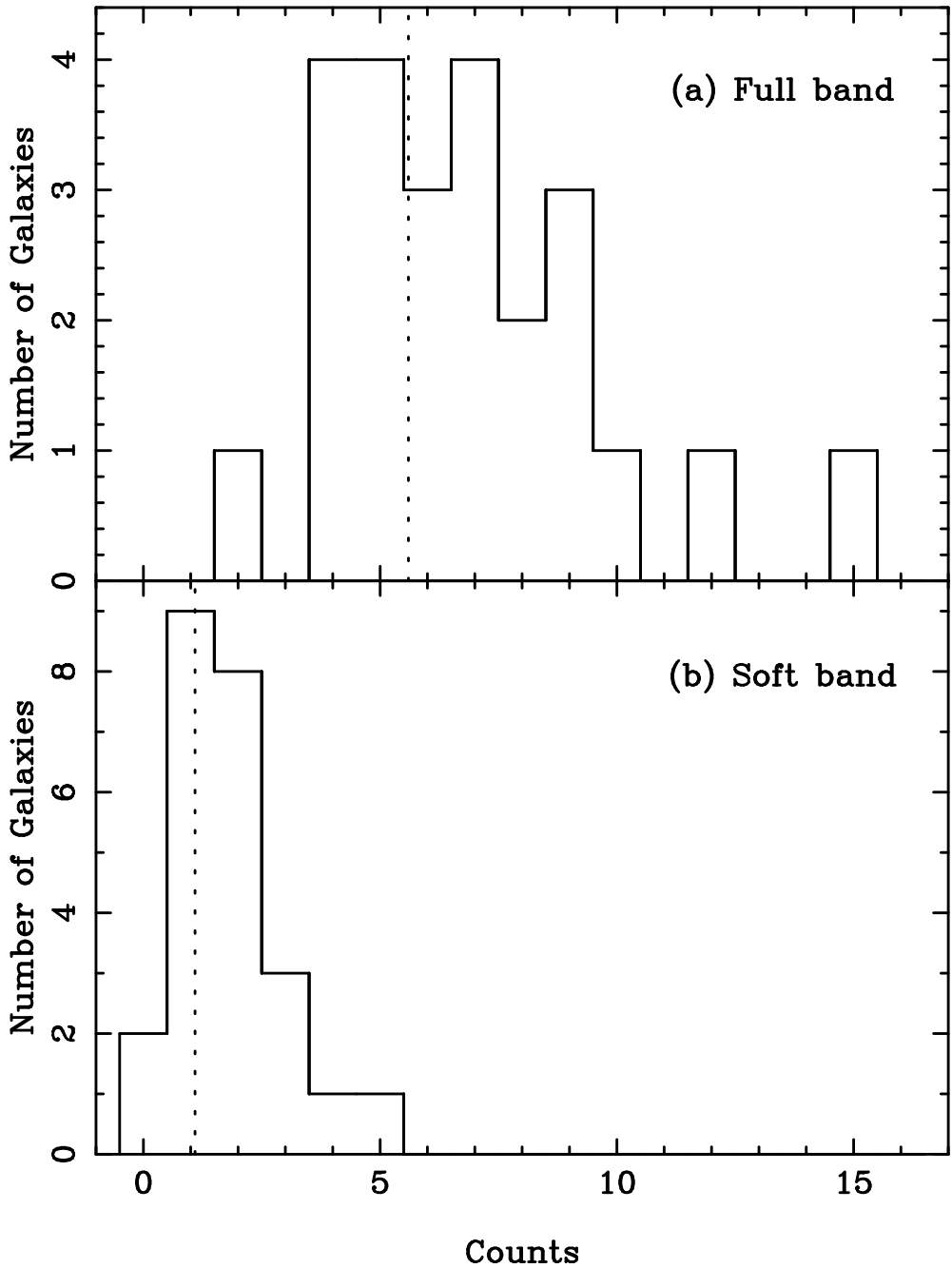}
\vspace{-0.1truein}
\caption{Histograms showing the number of (a) full-band counts and (b) soft-band 
counts obtained in the 30-pixel aperture for each of the galaxies used in the
stacking analyses. The average number of counts expected from background is 
5.6 for the full band and 1.1 for the soft band; these values are shown as the
vertical dotted lines.}
\end{figure}


\begin{figure}
\epsscale{1.0}
\figurenum{3}
\plottwo{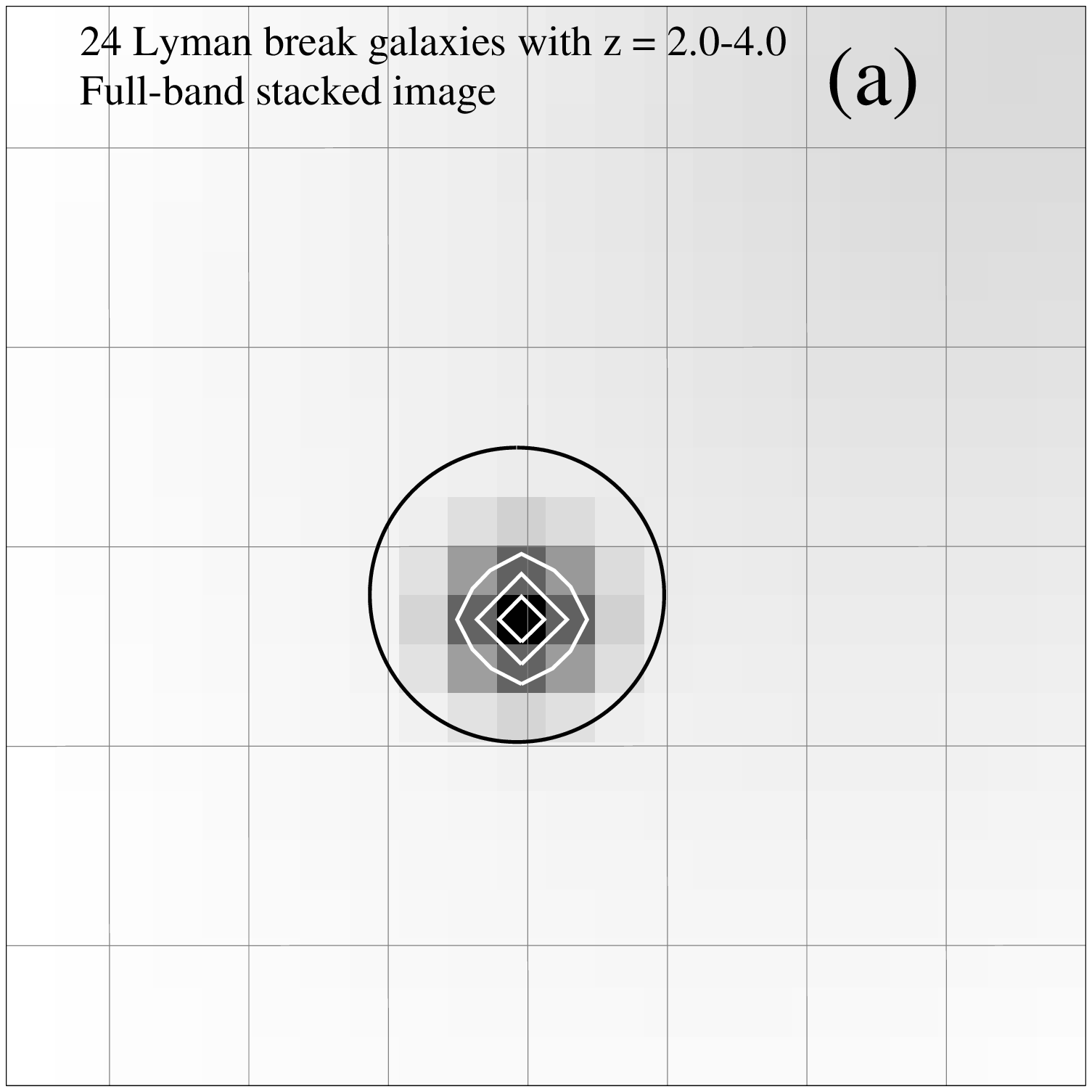}{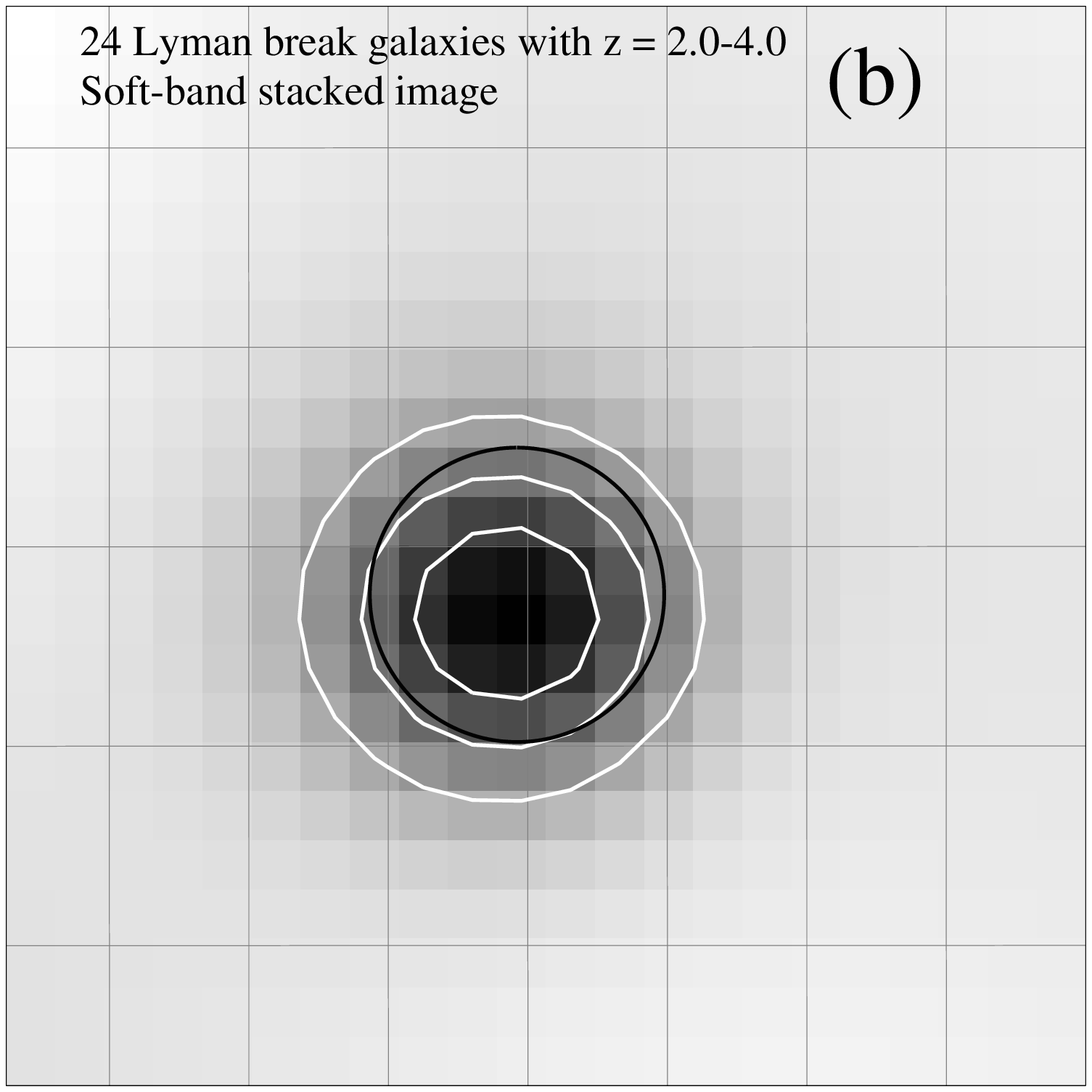}
\vspace{-0.1truein}
\caption{Stacked \chandra\ images of 24 Lyman break galaxies in the
(a) full band (22.44~Ms effective exposure) and (b) soft band
(22.46~Ms effective exposure). These images
have been made using the restricted ACIS grade set, and they have been
adaptively smoothed using the code of Ebeling, White, \& Rangarajan (2001). 
The images are $11\arcsec \times 11\arcsec$ in size, and each pixel is 
$0\farcs 5$ on a side. North is up, and East is to the left. The black
circles are centered on the stacking position and have a radius of 
$1\farcs 5$. The white contours are drawn at 85\%, 90\%, and 95\% of 
the observed maximum pixel intensity in each image.}
\end{figure}


\begin{figure}
\epsscale{0.7}
\figurenum{4}
\plotone{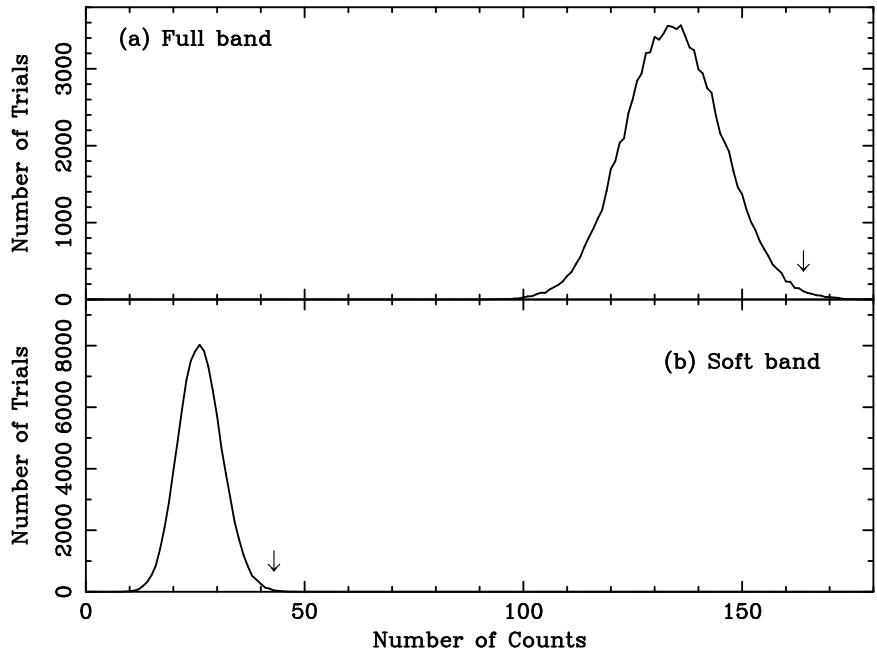}
\vspace{-0.1truein}
\caption{Results from Monte Carlo testing of the Lyman break galaxy stacking
analyses for the (a) full band and (b) soft band. Each panel shows the results
from 100,000 stacking trials plotted as the number of trials yielding 
a given number of counts. The resulting distributions are very nearly Gaussian. 
The arrow in each panel indicates the number of counts actually observed when the 
Lyman break galaxies are stacked.}
\end{figure}


\end{document}